\documentstyle[prd,aps,psfig]{revtex}
\begin{document} 
 
\tighten
\draft
\twocolumn[\hsize\textwidth\columnwidth\hsize\csname 
@twocolumnfalse\endcsname 

\title{Geometric Phase Transitions} 

\author{
Dorje C. Brody$^{*}$ 
and 
Adam Ritz$^{\dagger}$ 
} 
\address{$*$ DAMTP, Silver St., Cambridge CB3 9EW U.K.} 
\address{$\dagger$ Theoretical Physics Institute, University of 
Minnesota, 116 Church St., Minneapolis, MN 55455, U.S.A.} 

\date{\today} 

\maketitle 

\begin{abstract}
A model in statistical mechanics, characterised by the 
corresponding Gibbs measure, is a subset of the totality of 
probability distributions on the phase space. The shape 
of this subset, i.e., the geometry, then plays an important 
role in statistical analysis of the model. It is known that 
this subset has the structure of a manifold equipped 
with a Riemannian metric, given by the Fisher information 
matrix. Invariant quantities such as thermodynamic curvature 
have been studied extensively in the literature, although a 
satisfactory physical interpretation of the geometry has not 
hitherto been established. In this article, we investigate the 
thermodynamic curvature for one and two dimensional Ising 
models and report the existence of a geometric phase transition 
associated with a change in the signature of the curvature. 
This transition is of a continuous type, and exists for finite 
systems. The effect may be tested in principle by 
mesoscopic scale experiments. 
\end{abstract} 


\vskip2pc] 

In the area of parametric statistics, it has long been known 
that a useful and illuminating approach is to view statistical 
models as characterised by differentiable manifolds, equipped 
with the Fisher-Rao metric. Of particular interest in 
statistical inference is the notion of statistical divergence 
that measures how separated two probability distributions are, 
which is applied to study affinities amongst a given set of 
populations. It was discovered by Rao that this divergence can 
be measured by the geodesic distance determined from the Fisher 
information matrix \cite{burbea}. Analogous studies have been made 
in statistical mechanics, and numerous models including those 
exhibiting phase transitions were investigated \cite{geosm} (for 
a comprehensive list of references, see \cite{ruppeiner}). In 
particular, it was found that, for models exhibiting second order 
phase transitions, the curvature of the thermodynamic parameter 
space ${\cal M}$ diverges at the critical point, and that the 
scaling exponent of the curvature in the vicinity of the transition 
point is identical to that of the correlation 
volume \cite{ruppeiner}. Furthermore, 
for simple models such as the ideal gas, it has also been shown that 
the geodesic curves on ${\cal M}$ correspond to various equations 
of state for the system \cite{ingarden}. \par 

On the whole, most of the literature on the geometry of 
the thermodynamic parameter space has been devoted to analysing
properties in the thermodynamic limit, owing to the interest in 
critical phenomena. In this paper, however, we investigate finite 
size (volume) effects in the thermodynamic curvature for one and 
two dimensional Ising models. In particular, it is shown that, 
for a finite number ($N$) of spins, the phase diagram is clearly 
divided into two phases, namely, the positive and negative 
curvature phases. Furthermore, this phase separation disappears 
as $N\rightarrow\infty$. Because the size of the system is 
finite, the transition is continuous and is not accompanied by 
a singularity. Possible experimental realisations, based on the 
idea of generalised equations of state, are considered. \par 

Before turning to the analysis of finite size effects on the 
thermodynamic curvature for the Ising models, we first introduce 
the basic ideas behind the geometric approach to statistical 
mechanics. For this purpose, we consider the van der Waals gas 
model for vapour liquid transitions, because this is a 
nontrivial model that has nonetheless been studied quite 
extensively. \par 

In statistical mechanics, we begin by considering the 
probability distribution given by the Gibbs measure, 
\begin{equation} 
p(x|\theta)\ =\ \exp\left( -\sum_{i=1}^{r}\theta^{i}H_{i}(x) 
- \ln Z(\theta)\right)\ , 
\label{eq:gibbs} 
\end{equation} 
where $H_{i}(x)$ are Hamiltonian functions on the phase space, 
$Z(\theta)$ is the partition function, and $\{\theta^{i}\}$ are 
thermodynamic variables which may include inverse temperature, 
pressure, magnetic field, chemical potential, and so on. This 
density can be mapped to an element in a Hilbert space 
\cite{burbea} by the prescription 
$p(x|\theta)\rightarrow\psi_{\theta}(x)=\sqrt{p(x|\theta)}$. 
Then, because of the normalisation condition for $p(x|\theta)$, 
the square-integrable function $\psi_{\theta}(x)$ represents, 
for each fixed value of $\{\theta^i\}$, a point on the positive 
`octant' of the unit sphere ${\cal S}$ in Hilbert space. By 
continuously changing the values of $\{\theta^i\}$, this point 
clearly moves inside an $r$-dimensional subspace of ${\cal S}$. 
This subspace is the thermodynamic parameter space, or, for the 
above Gibbs measure, the maximum entropy manifold ${\cal M}$, and 
the metric on ${\cal M}$ induced by the underlying spherical 
geometry of ${\cal S}$ is called the Fisher-Rao metric. In 
particular, if we 
choose the parametrisation as given in (\ref{eq:gibbs}) and write 
$\partial_{i} = \partial/\partial\theta^{i}$, then this metric takes 
the simple form $G_{ij} = \partial_{i}\partial_{j}\ln Z$, from which 
the curvature can be computed via standard prescriptions in 
Riemannian geometry. \par 

Now suppose we consider the van der Waals model, where in equation 
(\ref{eq:gibbs}) we take $r=2$, $H_{1}(x)$ the energy, $H_{2}(x)$ 
the volume $V$, and $(\theta^{1},\theta^{2}) = (1/kT, P/kT)$. Thus, 
$Z(\theta)$ is a function of two variables, namely, temperature and 
pressure. In the case where ${\cal M}$ is two-dimensional, such 
as in this example and the Ising models to be considered later, the 
scalar curvature assumes a simple form  
\begin{equation} 
{\cal R}\ =\ -\frac{1}{2 G^{2}} 
\left| \begin{array}{lll} 
\partial^{2}_{1} \ln Z & \partial_{1}\partial_{2}\ln Z & 
\partial_{2}^{2}\ln Z \\ 
\partial^{3}_{1}\ln Z & \partial_{1}^{2}\partial_{2}\ln Z & 
\partial_{1}\partial_{2}^{2}\ln Z \\ 
\partial^{2}_{1}\partial_{2}\ln Z & 
\partial_{1}\partial_{2}^{2}\ln Z & \partial_{2}^{3}\ln Z  
\end{array} \right| \ , 
\label{eq:curvature} 
\end{equation} 
where $G={\rm det}(G_{ij})$. \par 

The phase space integral for the partition function is known to 
give rise to the cubic equation of state for the van der Waals 
model \cite{kuh}, which includes an essentially unphysical region 
in the parameter space where the pressure decreases as the volume 
is reduced. The conventional argument is to apply Maxwell's 
equal area rule to follow through the transition. This is 
sketched in Fig. 1. Note that not all of the region inside the 
Maxwell boundary is entirely unphysical, because in some part 
the pressure increases in decreasing volume. The unphysical 
region, on the other hand, is surrounded by the spinodal curve 
along which $\partial \theta^{2}/\partial v=0$, where 
$v=v(\theta^{1},\theta^{2})$ is the mean volume determined by 
the equation of state. This curve also contains the transition 
point where $\partial\theta^{1}/\partial v=0$. 

\begin{figure}[b] 
   \centerline{\psfig{file=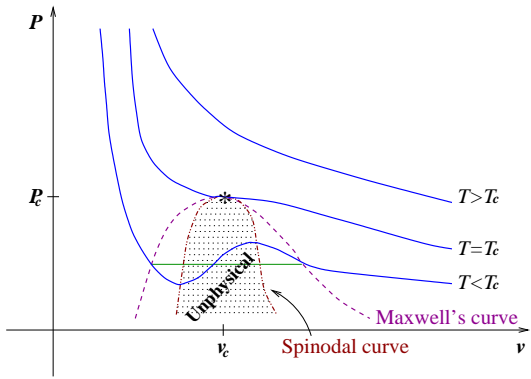,width=7cm,angle=0} }
\vskip .25cm
 \caption{The equation of state for the van der Waals model. 
The scalar curvature on the parameter space diverges along the 
spinodal boundary which envelopes the unphysical region. Thus, 
one can take the viewpoint that the curvature in some sense 
`prevents' the entrance to the unphysical domain.} 
\end{figure} 

For the van der Waals model, the scalar curvature ${\cal R}$ of 
the parameter space ${\cal M}$ has been calculated\cite{geosm}, 
and it was shown that the curvature diverges at the transition 
point, as well as along the spinodal curve. This intriguing picture 
can be further analysed by consideration of the surface ${\cal M}$ 
on the unit sphere ${\cal S}$. In the absence of a phase transition, 
${\cal M}$ is a single surface. However, below the transition 
temperature, that is, beyond the spinodal curve, the maximum 
entropy surface ${\cal M}$ proliferates into three surfaces,
associated with the three distinct roots of the equation of 
state. The scalar curvature, in particular, is singular along 
the curve where the proliferation occurs. Therefore, suppose we 
begin with a pure thermal state $|\psi(\theta)\rangle$ at high 
temperature, pure in the sense that it is a uniquely defined 
state in the Hilbert space. Then by reducing the temperature, this 
pure state `evolves' into a mixed state 
$\sum_{j}c_{j}|v_{j}\rangle \langle v_{j}|$ where $c_{j}^{2}$ 
determines the probability that the system volume is $v_{j}$. Then, 
by a suitable measurement to determine the volume of the system, 
this mixed state reduces to a pure state $|v_{k}\rangle$. This is 
illustrated in Fig. 2.\par 

\begin{figure}[b] 
   \centerline{\psfig{file=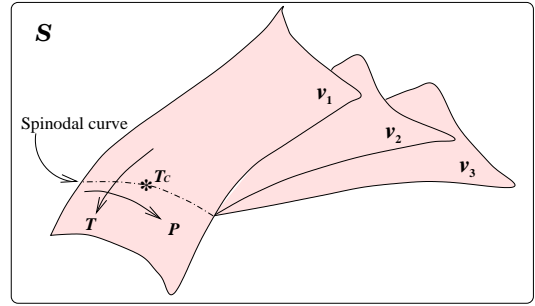,width=7cm,angle=0}} 
\vskip .25cm
 \caption{A schematic representation of the maximum entropy 
surface for the van der Waals 
model on the unit sphere ${\cal S}$ in Hilbert space. At high 
temperatures, the surface is uniquely defined, while at low 
temperatures, the surface is multi-valued, and can be labelled by 
the roots of the equation of state. Without further information, 
we cannot identify which one of the surfaces the equilibrium 
state would reach beyond the spinodal curve. 
} 
\end{figure} 

In the case of the van der Waals model, further physical arguments 
in fact rule out two of the unphysical roots for the equation of 
state, and we can single out one 
physical state $|v_{k}\rangle$. However, in a generic scenario of 
a phase transition, in the absence of any symmetry breaking field, 
the pure state characterising the equilibrium state turns into a 
mixed state through a geometric singularity, reflecting the 
multiplicity of the ground state. This is how the equilibrium 
theory for characterising symmetry breaking can be viewed in the 
geometric framework. Although the calculations involved here can 
be quite intricate, consideration of thermodynamic limit 
often leads to useful simplifications. \par 

While the theory of critical phenomena is of great interest in 
the approach outlined above, owing to various advances in 
experimental condensed matter physics, there is also a growing 
interest in understanding the effect of a finite system size on 
thermodynamic quantities. This motivates us to study, in particular, 
such effects on the thermodynamic curvature. This will be analysed 
here by consideration of the Ising models in one and two dimensions. 
Let us first study the one-dimensional Ising chain. The Hamiltonian 
for the system is given by 
\begin{equation} 
-\beta {\cal H}\ =\ \beta\sum_{i=1}^{N}\sigma_{i}\sigma_{i+1} 
+ h\sum_{i=1}^{N} \sigma_{i}\ , \label{eq:hamil} 
\end{equation} 
where $\{\sigma_{i}=\pm1\}$ are the spin variables, $\beta=1/(kT)$,
and the two-dimensional manifold ${\cal M}$ is parametrised by 
$(\theta^{1},\theta^{2})=(\beta,h)$. The components of the 
Fisher-Rao metric per spin are obtained by differentiating 
$\ln Z(\beta,h)/N$: 
\begin{equation} 
G_{ij} = \frac{1}{N}\partial_{i}\partial_{j} \left\{ \ln\left[ 
(\cosh h+\eta)^{N}+(\cosh h-\eta)^{N}\right]\right\}, 
\end{equation} 
where $\eta=\sqrt{\sinh^{2}h+e^{-4\beta}}$. Here, we set Boltzmann's 
constant $k=1$. If we compute the metric in the thermodynamic limit 
$N\rightarrow\infty$, then the resulting expression simplifies, and 
we obtain the thermodynamic curvature \cite{geosm}, given by 
${\cal R}=1 + \eta^{-1}\cosh h$, which is always positive. However, 
if we consider finite size effects, we observe that the curvature is 
no longer strictly positive. In particular, it was noted in 
\cite{dbar} that for a given point on parameter space, ${\cal R}$ 
could decrease and eventually become negative as $N$ was reduced. 
Indeed the transition to large negative values over a small range in 
$N$ is quite marked as one may observe in Fig.~3 where we consider 
the point $\{T=0.3,h=0.4\}$.
\begin{figure}[b] 
\centerline{%
   \psfig{file=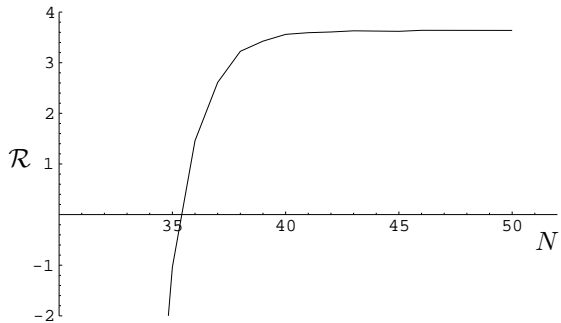,width=7cm,angle=0} 
   }%
\vspace{-2.5cm} \hspace{-0.0cm} ${\cal R}$

\vspace{0.7cm} \hspace{7cm} $N$

\vskip 1.2cm
 \caption{A plot of thermodynamic curvature ${\cal R}$ for the 
one-dimensional Ising model, with $\{T=0.3,h=0.4\}$. If the 
number ($N$) of spins is reduced from 35, the curvature rapidly 
grows to large negative values. When $N$ is increased, ${\cal R}$ 
approaches its thermodynamic limit ${\cal R}\sim 3.63$. } 
\end{figure} 

The result in Fig.~3 suggests the existence of a phase separation 
in terms of positive and negative curvatures. Indeed, because the 
components of the metric can be obtained explicitly for arbitrary 
finite $N$ \cite{dbar}, we can compute the curvature ${\cal R}$ as 
a function of $N$ and then numerically extrapolate the phase 
boundary along which the curvature vanishes. The result is shown 
in Fig.~4 for several values of $N$. Further numerical analysis 
shows that the parameter space is indeed separated into these 
two phases. Before seeking any physical consequences of this 
geometric phase separation, we would like to analyse the case of 
the two-dimensional Ising model, in order to indicate the 
universality of the phenomenon at least within the context of
ferromagnetic spin models. \par 

In the two-dimensional case, we also consider the Hamiltonian given 
in (\ref{eq:hamil}) parametrised by the inverse temperature and magnetic 
field, except the spin variables $\{\sigma_{i}\}$ are now defined 
on a toroidal lattice. For the two-dimensional Ising model with an 
applied field, an analytical expression for $\ln Z$ is not available 
for sufficiently large $N$. However, a Monte Carlo simulation turns 
out to be quite effective in analysing problems of this kind \cite{ejb}. 
In particular, if we return to the expression 
given in (\ref{eq:curvature}), we notice that each entry in the 
determinant is a combination of the products of central moments 
for terms in the Hamiltonian, and thus calculation of ${\cal R}$
is quite amenable to Monte Carlo analysis. \par

\begin{figure}[b] 
   \centerline{\psfig{file=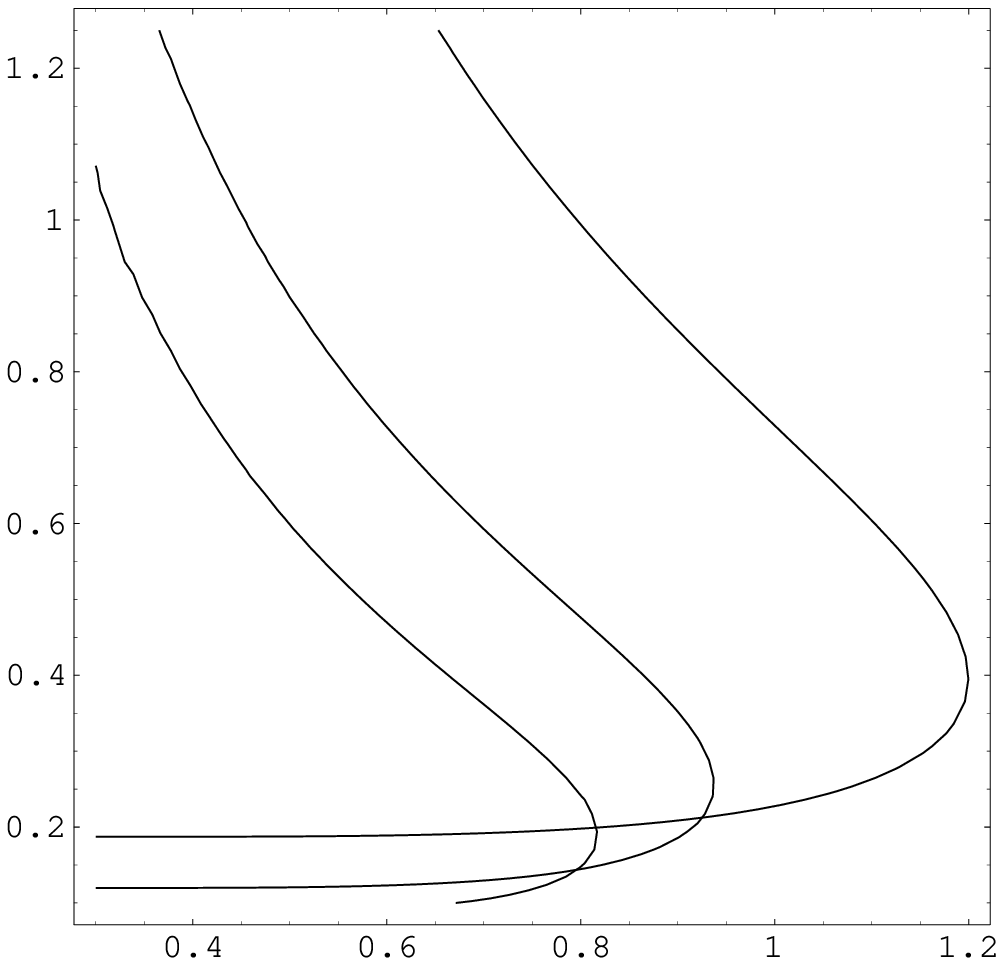,width=7cm,angle=0} }

\vspace{-4.0cm}\hspace{-0.2cm} $h$

\vspace{3.5cm}\hspace{4cm} $T$

\vspace{-6.2cm}\hspace{4.45cm} $N=8$

\vspace{1.2cm}\hspace{3.25cm} $N=12$

\vspace{1cm}\hspace{1.25cm} $N=16$

\vspace{3cm}
 \caption{The geometric phase diagram for the Ising chain in the 
$(T,h)$ plane. The curvature is positive in the high temperature 
phase, negative in the low temperature phase, and vanishes along 
the curves indicating the phase boundary for $N=8,~12,~16$. 
} 
\end{figure} 

We simulated the Ising spins on $8\times8$ and $16\times16$ lattices, 
using a standard Metropolis algorithm, and the results were combined 
from a series of runs using alternately sequential and random site 
updates, and hot and cold starts. The values of ${\cal R}$ were 
computed for a range of values of $(T,h)$, and the results clearly 
indicate the presence of a transition between positive and negative 
curvature regions. Numerical values for the curvature near this phase 
boundary, including statistical errors from a series of 10 runs with 
the differing initial conditions and update procedures mentioned above, 
are presented in Table~I. The relatively large errors are a result of 
significant cancellations between the third order cumulants used to 
construct the curvature. An interesting point to note is that the 
boundary is fairly stable for the two system sizes, in contrast to 
the shift 
observed in the 1D case. This is presumably related to the presence 
of the critical point at $(T_c\sim 2.27,h=0)$. Preliminary results 
for a similar analysis of the 3D Ising model also indicate a clear 
presence of a negative curvature region for low temperatures, 
although this region in parameter space is somewhat smaller than 
for the 2D case above. \par
 
In the foregoing analysis we have demonstrated the existence of 
positive and negative curvature phases for one and two dimensional 
Ising models. Whereas the discontinuities which are manifested in 
conventional phase transitions can only occur in the thermodynamic 
limit, these qualitative geometrical transitions occur even in the 
simplest finite-dimensional models.\par

The important issue we now address is whether there is any physical 
or observational consequence of this transition. While the answer 
appears to be yes insofar as the curvature is at least indirectly 
measurable, a more intriguing consequence arises once we recall 
the fact that the deviation of a collection of nearby geodesic 
curves is conditioned by the Riemann curvature tensor. In 
particular, if we consider a set of nearby geodesics in a region 
of ${\cal M}$, then these geodesics will tend to deviate apart if 
${\rm sgn}({\cal R})<0$, while the converse is true if 
${\rm sgn}({\cal R})>0$. Furthermore, the geodesics on ${\cal M}$ 
can be viewed \cite{ingarden} to play the role of generalised 
equations 
of state corresponding to different initial conditions. This can be 
seen from the fact that the leading term in the Taylor series 
expansion of the relative entropy is given by the infinitesimal 
line element $ds^{2}=G_{ij}d\theta^{i}d\theta^{j}$. In other words, 
we have $S(p(\theta)|p(\theta+d\theta))=\frac{1}{2}ds^{2}$ in the 
limit $d\theta\rightarrow0$. Hence we see that a geodesic joining 
two nearby points on ${\cal M}$ corresponds to a trajectory 
that extremises the entropy, and thus has a natural parallel 
with an equation of state.\par 

On the other hand, for any given point in ${\cal M}$ there are 
infinitely many geodesic curves that pass this point, each of which 
is subject to an initial condition, which may include the 
requirement of an adiabatic or isothermal change of the system. In 
this case, two nearby equations of state would deviate away if the 
sign of ${\cal R}$ is negative. Therefore, an experimental 
verification of the phase separation requires an extensive 
analysis of equations of state for various different processes. 
Given such data it may be possible to identify qualitatively 
different behaviours for the two phases.\par 

\begin{table}[t]

\begin{tabular}{||c||c|c|c|c|c||}
  $8\times8$ &    $h$=0.02  & 0.18    & 0.34    & 0.50     & 0.66 \\ \hline
 $T$=2.2     &  -2.8   &  -9.2   & -4.5    & -5.8    & -0.1 \\ 
  2.3        &  -1.9   &  -5.9   & -4.7    & -2.5    & +1.5 \\
  2.4        &  +0.6   &  -2.8   & -1.2    & -1.4    & +0.1 \\
  2.5        &  +1.1   &  -2.8   & -1.1    & -1.3    & +0.7 \\
  2.6        &  +2.8   &  -2.4   & -1.3    & -0.5    & +3.3 \\ 
  2.7        &  +4.1   &  -2.5   & -0.7    & -0.2    & +0.2 \\
  2.8        &  +5.6   &  -1.2   & -0.1    & +0.1    & +1.1 \\
  2.9        &  +7.0   &  -0.7   & +0.7    & +1.0    & +2.0 \\
  3.0        &  +7.3   &  +2.4   & +0.1    & +1.5    & +3.0
\end{tabular}

\begin{tabular}{||c||c|c|c|c|c||}
  $16\times16$     & $h$=0.02 & 0.18   & 0.34     & 0.50    & 0.66 \\ \hline
 $T$=2.2     & -58.3    & -7.8  & -5.7  & -2.8  & -2.5 \\
   2.3       & -55.1    & -7.1  & -5.6  & -3.5  & -2.6 \\
   2.4       & -32.4    & -6.4  & -5.3  & -2.4  & -2.1 \\
   2.5       & -6.9     & -2.8  & -2.6  & -1.2  & +0.2 \\
   2.6       & +13.2    & -2.1  & -0.7  & -0.9  & +1.1 \\
   2.7       & +22.1    & +0.9  & -0.8  & -0.1  & +1.6 \\
   2.8       & +24.3    & +1.8  & -1.0  & -0.1  & +2.5 \\
   2.9       & +25.8    & +2.4  & +0.6  & +0.1  & +2.6 \\
   3.0       & +23.9    & +2.9  & +0.7  & +0.3  & +1.3 
\end{tabular} 
\vskip .25cm
 \caption{The values of the curvature ${\cal R}$ $(\pm 0.6)$ of the 
parameter space for the two-dimensional $8\times8$ and $16\times16$ 
Ising lattice. The results clearly indicate the phase separation, 
with a negative curvature phase for low temperatures.} 
\end{table}


In summary, we have shown the existence of phase separations in 
terms of the signature of the thermodynamic curvature for finite 
Ising spins in one and two dimensions. Owing to the fact that 
${\cal R}$ is expressible in terms of correlation functions of 
order $\leq6$ ($\leq3$ in the present case), a Monte Carlo 
computation is straightforward, in contrast to the renormalisation 
group approach which generally requires knowledge of an infinite 
number of correlation functions. Therefore, it would be of interest 
also to analyse the curvature for lattice gauge theories in a finite 
volume. Finally, we have argued that geodesics on ${\cal M}$ 
generalise the notion of an equation of state, for arbitrary initial 
conditions, and thus experimental realisation may be obtained in 
principle by the analysis of equations of state for these systems at 
mesoscopic scales. \par

Another interesting open question in this regard is the direct 
physical interpretation of the curvature ${\cal R}$. We note that 
${\cal R}$ is determined by the parallel transport of tangent 
vectors over ${\cal M}$. In particular, by following a closed 
contour ${\cal C}$ on ${\cal M}$, a tangent vector is rotated by an 
angle given by the integral of ${\cal R}$ over the area inside 
${\cal C}$ (holonomy). In ordinary thermodynamics, Maxwell's 
relations guarantee that any closed contour ${\cal C}$ is 
thermodynamically trivial, provided ${\cal C}$ does not surround a 
critical point or straddle the spinodal curve. Therefore, if there 
is a physical interpretation of tangent vectors over the maximum 
entropy surfaces, then the geometric notion of holonomy leads to a 
new thermodynamic property whereby a system does not return to its 
original state after going around a contour, a concept analogous 
to magnetic hysteresis.

DCB is supported by Particle Physics and Astronomy Research Council. 
The authors acknowledge N. Rivier for stimulating discussions. 

$*$ d.brody@damtp.cam.ac.uk
 
$\dagger$ aritz@mnhepw.hep.umn.edu

\end{document}